\documentstyle[11pt,newpasp,twoside,epsf]{article}
\begin{document}

\title{Faint Quasar Surveys}
\author{Patrick B. Hall}
\affil{Department of Astronomy, University of Toronto, 60 St. George Street,
Toronto, Ontario M5S 3H8, Canada}

\begin{abstract}
Faint quasar surveys are necessary complements to bright quasar surveys
to remove the degeneracy between redshift and luminosity 
inherent in any single magnitude-limited sample.
I discuss two ongoing surveys for faint quasars at $3.3 < z < 5$ and 
$5 < z < 6$ using imaging data from the Big Throughput Camera on the CTIO 4-m.
I also discuss a sample of faint spectroscopically selected AGN with $z < 4.7$
found serendipitously in the CNOC2 Field Galaxy Redshift Survey.  Faint quasars
at $2 < z < 3.5$ from this sample, when combined with literature data on more 
luminous quasars at the same $z$, show evidence for a much weaker Baldwin Effect
in C\,{\sc iv}\,$\lambda$1549 and Ly$\alpha$ than previously seen at $z < 1.5$.
This may imply that the slopes of the Baldwin Effect for these transitions 
evolve with redshift, steepening with cosmic time.
Finally I discuss the prospects for extending faint quasar surveys to 
$z \sim 7$ using near-IR followup of very red
objects in the Red-Sequence Cluster Survey.
\end{abstract}

\section{Introduction}

Active Galactic Nuclei (AGN) 
are very luminous objects whose emission
is thought to be powered by the release of gravitational potential energy
from matter falling into supermassive black holes.
Most AGN physics beyond that simple statement
are a matter of dispute at some level, as are many of the physical differences
underlying the observed differences in the spectra of AGN.

One reason for at least part of our lack of understanding
may be selection effects.  Most high-luminosity AGN --- quasars, with $M_B<-23$
--- have been selected using rest-frame ultraviolet-optical colors.
Thus dust extinction has probably masked some portion of the dominant 
radio-quiet quasar (RQQ) population from detection (Webster et al. 1995).
Important steps toward eliminating this bias are beginning to appear from
near-IR selected samples (e.g. from 2MASS; see Cutri et al., this volume) 
and from X-ray selected samples 
(Kim \& Elvis 1999).
Radio selection may not sample as distinct a population of AGN as once thought,
given the lack of bimodality in $L_{rad}/L_{opt}$ between RQQs and RLQs seen in
the FIRST Bright Quasar Survey (White et al. 2000), 
but much deeper radio surveys 
would be needed to select the majority of radio-quiet AGN at redshifts $z < 5$.

Another important selection effect is that
any single flux-limited sample of AGN (even a near-IR or X-ray selected one)
will exclude high-redshift, low-luminosity objects.  This leads to a
degeneracy between the redshift and luminosity dependences of any AGN property.
Thus, bright quasar surveys need to be accompanied by faint quasar surveys
to fill in the $L$$-$$z$ plane and remove the redshift--luminosity degeneracy,
and to define the full quasar luminosity function.

At the moment optical color selection of AGN is 
the method being used to assemble the largest samples of AGN, namely
2.5$\times$10$^4$ objects from the 2dF Quasar Survey (2QZ; Boyle et al., this 
volume) and 10$^5$ from the Sloan Digital Sky Survey (SDSS; York et al. 2000).
Moderately deep wide-field optical imaging can provide the data needed for
complementary faint color-selected quasar surveys.
Several surveys are underway which, despite their small size, will form
useful comparisons to the 2dF and SDSS surveys (AGN will also turn up in faint
galaxy surveys such as discussed by Le Fevre et al., this volume).
The eventual goal for optical quasar studies is to reach $M_B=-23$
at all redshifts $z<7$. 

\section{Two Surveys for Faint Quasars at $3.3 < z < 6$}

The Big Throughput Camera on the CTIO 4-m (Wittman et al. 1998)
offered the first opportunity
to image the wide fields required for faint quasar surveys at $z>3$.
A group of quasar enthusiasts --- Julia Kennefick (Caltech), Pat Osmer
\& Eric Monier (Ohio State), Malcolm Smith (CTIO), Richard Green (NOAO) and
myself --- have used the BTC for two separate but complementary surveys.

In 1997--1999, we obtained 7 deg$^2$ of $BRI$ imaging data for a 
Big Faint Quasar Survey.
We expect to find $\sim$120 quasars at $3.3 < z < 4$ 
and $\sim$40 at $4 < z < 5$ via followup spectroscopy to $R$=23.5.
At these redshifts quasars have redder $B-R$ and bluer $R-I$ colors
than the stellar locus.
This survey will define the shape of the quasar luminosity function 
from $-26.5 < M_B < -23.5$ at $z > 3.3$, and its evolution at $3 < z < 5$,
with accuracy equal to current knowledge for $M_B < -26.5$
(Fan et al. 2000).
A data paper is nearing submission (Hall et al., in preparation),
and we hope to begin spectroscopy in earnest in 2001.

A second survey was also begun in 1997 to look for $z$$>$5
quasars using $VI$ imaging, dubbed BTC40 for its coverage of 40 deg$^2$ 
(Kennefick et al., in preparation).
A selection criterion of $V-I>3$ 
is sensitive to $z>5$ quasars and only the latest M dwarfs.
However, the surface density of these very late M dwarfs is
sufficiently high that they dominate candidates selected this way.  
Thus we added $z'$ band imaging to the survey to have two colors with which to
distinguish $z>5$ quasars from stars.
We expected anywhere from 1 to 25 $z>5$ quasars to $I$=22 in the survey.
To date we have spectroscopically confirmed two $4<z<5$ quasars,
but none at $z>5$.
Part of this is the difficulty of following up $I>21$ candidates on 
4-m class telescopes.
Our $z>4.5$ quasar surface density agrees with that found in the SDSS 
(Fan et al. 2000), but we hope to extend our spectroscopy to lower
luminosities using 8-m class telescopes in the near future.

\section{Faint Quasars in the CNOC2 Field Galaxy Redshift Survey}

\begin{figure}
\plottwo{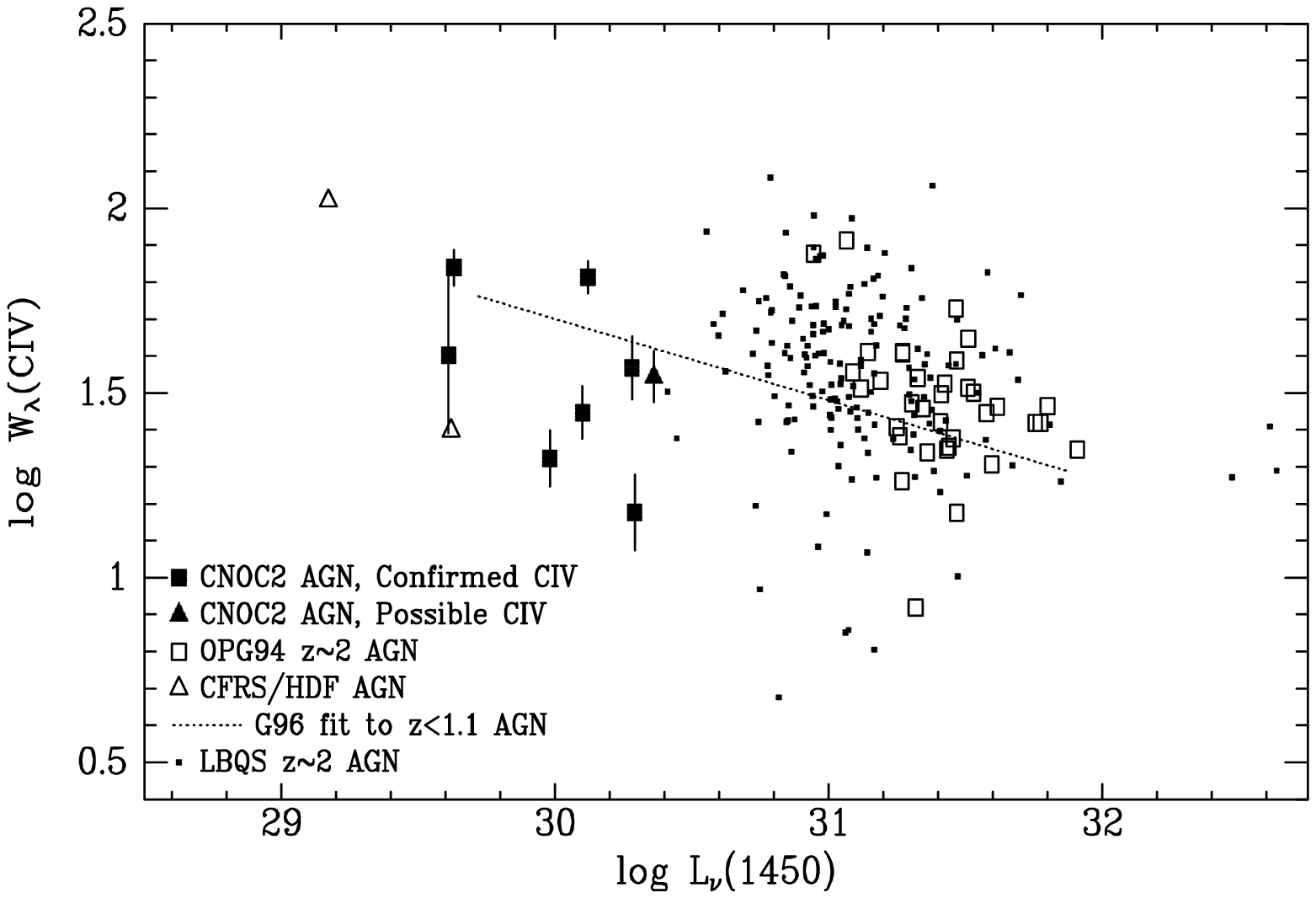}{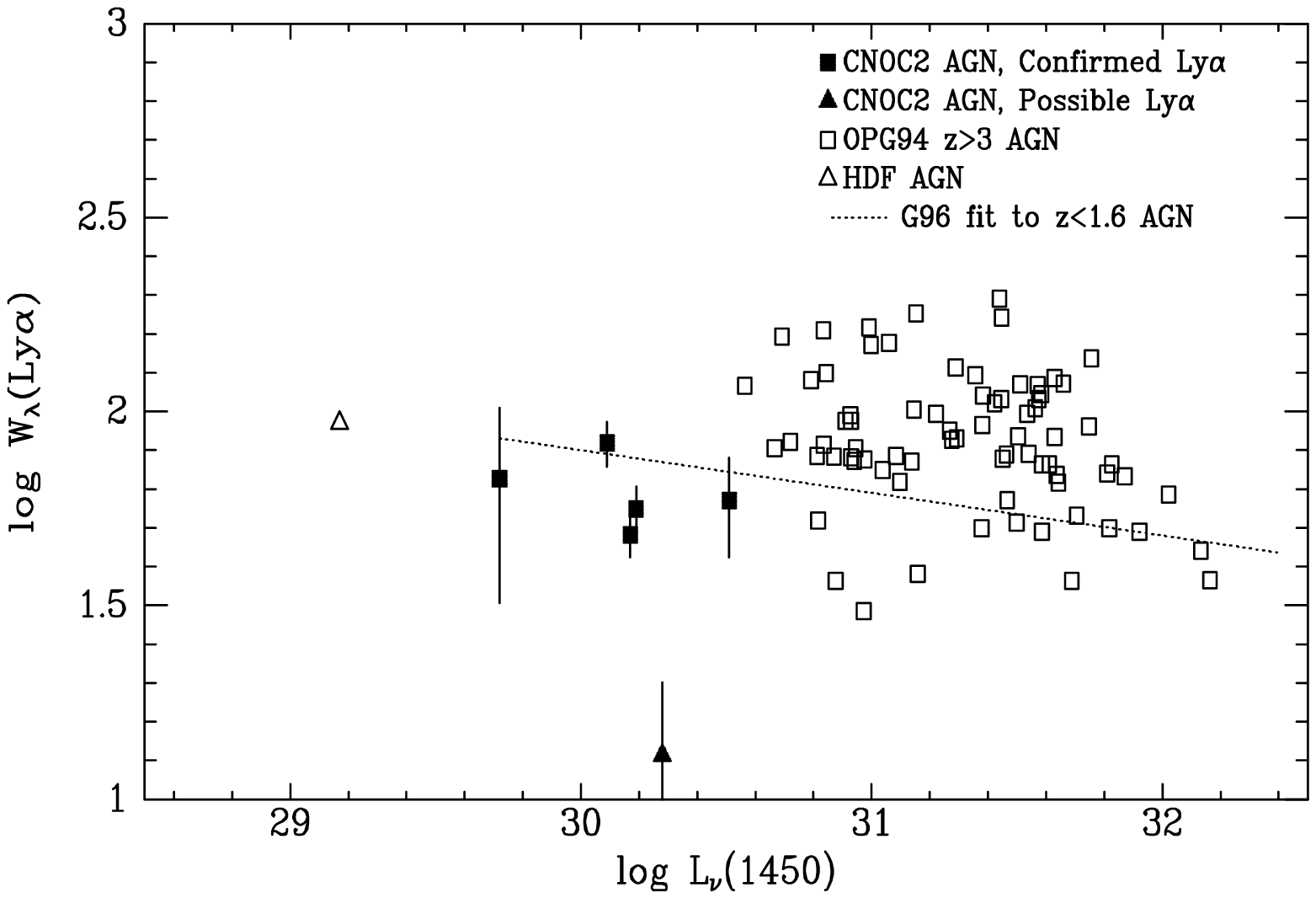}
\caption{The Baldwin Effect for C\,{\sc iv} at $z\sim2$ (left) and Ly$\alpha$ 
at $z\sim4$ (right).
Filled squares and triangles are CNOC2 AGN with firm and tentative redshifts,
respectively;
points are LBQS AGN;
open squares are AGN from Osmer et al. (1994);
and open triangles are AGN from the CFRS and HDF.
The dotted line shows the correlation found by Green (1996) for $z<1.6$ AGN
over the range of 1450\AA\ continuum luminosities in his sample;
it is clearly a poor fit for AGN at $z>2$.
}
\end{figure}

The Canadian Network for Observational Cosmology Field Galaxy Redshift Survey
(CNOC2) obtained spectra for about 6200 galaxies to a nominal limit of $R$=21.5.
As a side benefit it yielded a 
sample of at least
47 AGN with $0.27 < z < 4.67$ and average $M_B = -22.25$ (Hall et al. 2000),
spectroscopically selected by broad emission lines, Ne\,{\sc v} emission,
or Fe\,{\sc ii} absorption.
%
At least 80\% of the CNOC2 AGN can be recovered by color selection, 
in agreement with previous, mostly smaller 
samples for which multiple selection methods were employed.
This does not address the ability of optical flux-limited samples
to recover AGN which experience significant dust extinction, just the fraction
of AGN to a given optical flux limit
which have colors sufficiently distinct from stars to be identified.
Note that $\sim$20\% of these AGN are classified as resolved or
probably resolved (in CFHT seeing),
a potential selection effect which must be considered
in faint quasar surveys which target only unresolved objects.

The CNOC2 AGN include several unusual objects:
one with a very strong double-peaked Mg\,{\sc ii} emission line,
one with the first recognized O\,{\sc iii}\,$\lambda$3133 broad absorption line,
and one with as yet unidentified emission and absorption lines.  
The broad emission line subsample of predominantly low-luminosity AGN, 
although small, has some intriguing properties.
It may have a higher incidence of associated Mg\,{\sc ii} and C\,{\sc iv}
absorption, possibly because such absorption is anti-correlated
with optical luminosity or is becoming less frequent with cosmic time,
or possibly because the sample is not biased against objects
	reddened by dust associated with the absorbing gas or
with resolved spatial structure such as companions, tidal tails or
large host galaxies, environments which may be more likely to have
extensive gas envelopes which show up as associated absorption.

The quite similar spectra of broad-line AGN over a range of $\sim$10$^6$ in
luminosity implies that the 
parameters determining the emergent spectra scale nearly uniformly with
luminosity.
For a given line,
an anticorrelation of emission line equivalent width with continuum luminosity
--- a Baldwin Effect (BEff) --- 
means that the scaling is not exactly homologous for that line.
The broad emission line CNOC2 AGN subsample has average equivalent widths 
($W_{\lambda}$)
for Mg\,{\sc ii}\,$\lambda$2798 and C\,{\sc iii}]\,$\lambda$1909 which
agree with the predictions of previous studies of the BEff, but as seen
in Figure 1, the average $W_{\lambda}$ for C\,{\sc iv}\,$\lambda$1549
and Ly$\alpha$ are
smaller than predicted by previous studies of the BEff at lower redshift.
%
This may imply that the slopes of the C\,{\sc iv} and Ly$\alpha$ Baldwin
effects evolve with redshift, steepening with cosmic time.
Other faint AGN samples are consistent with these results,
namely AGN from the CFRS and HDF (see Figure 1),
spectroscopically selected AGN from the CNOC1 
Cluster Galaxy Redshift Survey, 
and low S/N spectra of AGN from the Deep Multicolor Survey 
(Osmer et al. 1998). 
A similar result was also reported by Zitelli et al. (1992) for C\,{\sc iv},
and by Boyle et al. (this conference) for more luminous 2QZ quasars.

A current favored model explains the BEff as a result of the luminosity
dependence of AGN continuum spectral energy distributions (SEDs), 
wherein more luminous AGN have softer ionizing continua (Osmer \& Shields 1999).
Evolution in the BEff might indicate evolution with cosmic time
of AGN SEDs at a given luminosity, which might arise from evolution in the
relation between AGN luminosity and black hole mass (Wandel 1999).
However, an earlier model for the C\,{\sc iv} BEff postulated a smaller
ionization parameter $\Gamma$ (the ratio of ionizing photon to gas densities) 
in more luminous AGN and predicted 
no Ly$\alpha$ BEff 
(Shields \& Ferland 1993).
A $\Gamma$-$L$ anticorrelation has also
been recently suggested on entirely independent grounds, namely to explain
reverberation mapping results on a subsample of Palomar-Green AGN
(Kaspi et al. 2000).
Given the theoretical implications, further data on high-$z$ low-luminosity AGN
from well-understood samples is needed to confirm our possible evidence for
redshift evolution of the slope of the Baldwin Effect in Ly$\alpha$ and 
C\,{\sc iv}.

\section{A Faint $z>5.5$ Quasar Search in the Red-Sequence Cluster Survey}

The $BRI$ and $VIz'$ color selection techniques used to identify
$3.3<z<5$ and $5<z<6$ quasars, respectively,
can be extended to $5.5<z<7$ quasars using $Rz'J$ imaging.
We are conducting such a survey using data from the Red-Sequence Cluster Survey
(RCS; Gladders \& Yee 2000 and Gladders, this conference),
a 100 deg$^2$ $Rz'$ survey for $z<1.4$ galaxy clusters.
As a side benefit, this survey should yield a sample of several hundred
T, L and M7--M9 dwarfs.

Candidate $5.5<z<7$ quasars and brown dwarfs are identified to
$z'=21.6$ 
by requiring $r'-z'\geq4$; 
the only plausible contaminant 
at such colors is extremely luminous $z>5$ galaxies.
$J$-band followup then separates the three populations:
T dwarfs have red $z-J$, L dwarfs have moderate $z-J$,
and $z>5.5$ quasars have blue $z-J$ (Figure 2 and Zheng et al. 2000).
In pilot $J$ imaging of 28 
targets sparsely sampled from $\sim4$ deg$^2$,
we are 28 for 28 in confirming that our $r'-z'>4$
candidates are real and fall within the expected $z-J$ color ranges:
23 have colors of M7--L8 dwarfs and 5 of L8 to T dwarfs (Figure 2).

%

The RCS can reach 2--2.5 magnitudes fainter than SDSS or 2MASS
in its selection of candidate $z>5.5$ quasars and L and T dwarfs.  
This means fainter quasars (and more of them)
and cooler brown dwarfs (on average).
The full 100 deg$^2$ survey could yield anywhere from 0 to $\sim$100 $z>5.5$
quasars,
as well as more than 50 T dwarfs and 450 L dwarfs (D'Antona et al. 1999).

\begin{figure}
\plotone{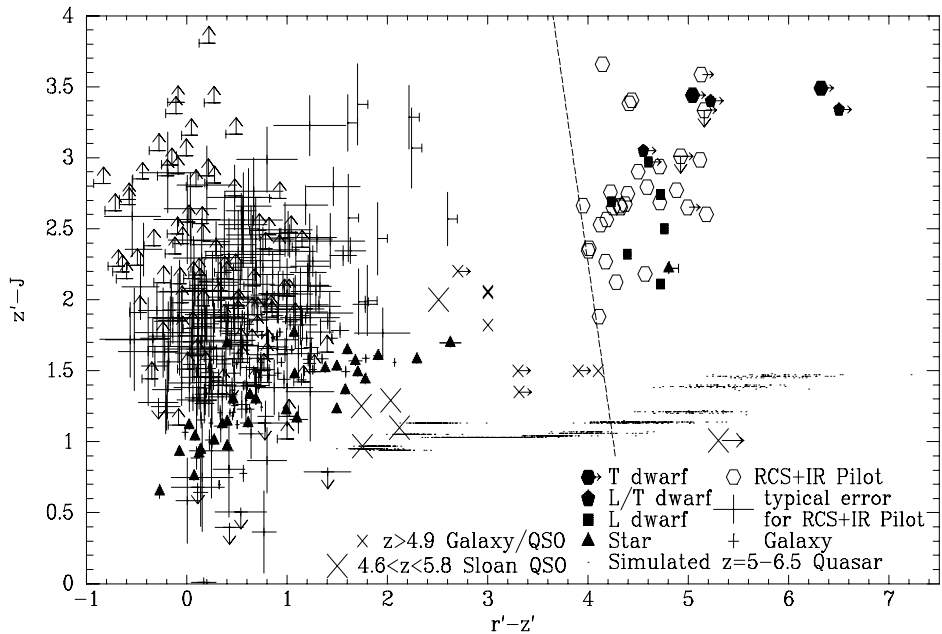}
\caption{$r'$$-$$z'$/$z'$$-$$J$ color-color diagram, with a dashed line showing
our {\bf selection criterion} $r'$$-$$z'$$>$4 
for $z>5.5$ quasars and brown dwarfs in the RCS.
%
The various populations are, from left to right:
Morphologically classified {\bf Galaxies} (error bars) and
{\bf main sequence Stars} (filled triangles with error bars)
from 17~arcmin$^2$ of $r'z'JK$ imaging to $K$$\sim$21
around two $z$$\sim$1.5 quasars (Hall et al. 1998).
Upper/lower limits are indicated with arrows or error bars with tick marks.
Most objects in deep imaging samples are much bluer than our selection criteria.
{\bf $z$$>$4.9 Galaxies/QSOs} (small crosses) from the literature are plotted
at their measured or estimated colors (Hu et al. 2000).
Our color criterion might select some $z$$>$4.9 galaxies, but any in our
$z'$$\leq$21.6 sample would be extremely luminous and well worth studying.
{\bf Simulated Quasars} (points) at $z$=5 ($r'$$-$$z'$$\sim$2)
to $z$=6.5 ($r'$$-$$z'$$\sim$6) lie below the stellar and
brown dwarf loci, even when accounting for the $\pm$0\fm5 scatter around
the points expected from the dispersion in quasar spectral properties.
Of the six {\bf 4.6$<$$z$$<$5.8 Sloan QSOs}
(large crosses; Fan et al. 2000 \& Zheng et al. 2000),
we would have selected the one at $z$=5.8 ($r'$$-$$z'$$\geq$5.3)
but none of the five at 4.6$<$$z$$<$5.3 ($r'$$-$$z'$$\sim$2), as expected.
{\bf Confirmed L dwarfs} (filled squares; Fan et al. 2000)
define the brown dwarf locus from M7 (lowest filled square,
$z'$$-$$J$$\sim$2) to L8 (highest filled square, $z'$$-$$J$$\sim$3).
{\bf Confirmed L/T dwarfs} (filled pentagons; Leggett et al. 2000) and
{\bf Confirmed T dwarfs} (filled hexagons; Strauss et al. 1999 \& Tsvetanov
et al. 2000) show that objects L8 and later have $z'$$-$$J$$\geq$3.
{\em The separation of T dwarfs, L dwarfs, and $z$$>$5.5 quasars from each
other and from typical stars and galaxies is obvious.}
{\bf RCS targets} (open hexagons) sparsely sampled from 4~deg$^2$ of data
and with IR pilot observations show the efficiency of our selection criteria.
23 of 28 observed are candidate M7 to L8 dwarfs,
while the 5 reddest in $z'-J$ are candidate L8 to T dwarfs.
}
\end{figure}

\section{Conclusions}

Large surveys of bright quasars need complementary 
surveys of faint quasars if redshift and luminosity dependences
of various quasar properties are to be disentangled.
Together these surveys will enable study of the properties and evolution of
quasars 
over the entire optically accessible $z<7$ range.

The BFQS and BTC40 surveys are first steps toward large faint quasar surveys,
for which 4-m class telescopes can provide imaging and some spectroscopy, 
but for which 8-m class telescopes are needed 
for fainter candidates and 
detailed spectral studies.
%
The RCS is an excellent database in which to look for low-luminosity
$5.5 < z < 7$ quasars, using snapshot near-IR imaging to separate them from 
very late M, L and T dwarfs with similarly red $R-z'$ colors.

Comparison of low-luminosity quasars from the CNOC2 survey to high-luminosity
quasars at the same redshifts suggests an unexpected variation with redshift
of the Baldwin Effect for C\,{\sc iv} and Ly$\alpha$,
illustrating the potential for surprises 
in the new era of wide-field astronomy.

\acknowledgments
I thank all my collaborators for permission to discuss the results
obtained by the teams to which I belong.

\end{document}